\documentclass{article}
\usepackage{fleqn}
\usepackage{amsmath,amssymb}

\begin{document}

\begin{center}
{\bf\Large The cosmological evolution  of the cosmological plasma with interpartial scalar interaction.\\[12pt] I. The canonic formulation of classical scalar interaction}. \\[12pt]
Yu.G. Ignat'ev\\
N.I. Lobachevsky Institute of Mathematics and Mechanics, Kazan Federal University, \\ Kremleovskaya str., 35, Kazan, 420008, Russia
\end{center}

\begin{abstract} On the basis of Hamilton a formalism the dynamic equations of movement scalar  charged particles in a classical scalar field are formulated. Unlike earlier published works of the author the model with zero own weight of particles is considered. Linear integrals of movement are found and ambiguity of communication between kinematic speed and an impulse of particles is specified.
\end{abstract}

\section{Introduction}
Recently the considerable quantity of works on late acceleration of the Universe is published. For the solution of secondary acceleration of the Universe problem in many works it is offered to change fundamental principles of physics radically. However, now there are some arguments in favour of that difficult, multicomponent, classical physical systems also can result in secondary acceleration of the Universe. In particular, such arguments have been resulted by D. Galtsovym, and also the author of article in reports at seminar Gracos-2009 (see, for example, [1], [2]). In the quoted work [2] the example cosmological evolutions completely degenerate Fermi-system with scalar interaction of particles, with initial inflation and late acceleration has been resulted. In V. Zhuravlyov's work [3] the cosmological evolution of the two-componental system consisting of an ideal liquid and a scalar field was investigated. In these works it is shown, that such cosmological models can have an initial inflationary stage and later acceleration. Thereby, the cosmological models with a multicomponent matter in a condition to describe the basic observant data about Universe expansion. Some instructions on possibility of such behaviour of difficult systems with scalar interaction of particles have been given also in works [4], [5]. Unlike two-componental system «a scalar field + an ideal liquid» in which interaction of components is carried out only through gravitation, we will consider statistical systems of the scalar charged particles in which some sotrs of particles can to directly interact with a scalar field through some fundamental\textit{ scalar charge}. On the other hand, the statistical system, possessing, generally speaking, a nonzero scalar charge and itself being a source of a scalar field, can effectively influence a scalar field, operating its behaviour. Such scalar interaction has been entered in the general relativistic kinetic theory in 1983 by the author of article [6, 7, 8, 9] and by and by -- G.G. Ivanov [10]. In particular, in works [7, 8] on the basis of the kinetic theory the self-consistent system of the equations describing statistical system of particles with scalar interaction is received.

\section{Dynamic of the scalar interaction particles}
\subsection{The canonical equations of motion}
The canonical equations of movement of a relativistic particle relative pair canonic conjugate dynamic variables $x^{i} $  (coordinates) and $P_{i} $  (the generalized momentum) has the form (see, for example, [7]):
\begin{equation} \label{GrindEQ__1_}
\frac{dx^{i} }{ds} =\frac{\partial H}{\partial P_{i} } ;\quad \quad \frac{dP_{i} }{ds} =-\frac{\partial H}{\partial x^{i} } ,
\end{equation}
 where $H(x,P)$ is the relatovistic invariance Hamilton function. Obtainig the total derivative of the function of the dynamical variables $\Psi (x^{i} ,P_{k} )$, taking account \eqref{GrindEQ__1_}, we find:

\begin{equation} \label{GrindEQ__2_}
\frac{d\Psi }{ds} =[H,\Psi ],
\end{equation}
 where the invariance Poisson's brakets are intoduced:

\begin{equation} \label{GrindEQ__3_}
[H,\Psi ]=\frac{\partial H}{\partial P_{i} } \frac{\partial \Psi }{\partial x^{i} } -\frac{\partial H}{\partial x^{i} } \frac{\partial \Psi }{\partial P_{i} } \; .
\end{equation}
In consequence of \eqref{GrindEQ__3_} Hamilton function  is the movement integral of a particle, - this movement integral is called as rest-mass of a particle (we will notice, that here and further the universal system of units $G=c=\hbar =1$ is accepted):

\begin{equation} \label{GrindEQ__4_}
\frac{dH}{ds} =[H,H]=0,\Rightarrow H=\frac{1}{2} m^{2} ={\kern 1pt} Const{\kern 1pt} .
\end{equation}
The relation \eqref{GrindEQ__4_} is called as a normalisation relation.  The invariant Hamilton function is defined ambiguously. Really, in consequence of \eqref{GrindEQ__3_}, if $H(x,P)$ there is a function of Hamilton, also any continuously differentiated function $f(H)$ also is function of Hamilton. In work of the author [6] relativistic- invariant Hamilton function of particles with the scalar charge $q$, being in a scalar field with potential $\Phi $ was introduced by relation:
\begin{equation} \label{GrindEQ__5_}
H(x,P)=\frac{1}{2} m\left[\frac{(P,P)}{m+q\Phi } -q\Phi \right],
\end{equation}
where $(a,b)$ here and further there is a scalar product of vectors of  four-dimensional vectors $a$ and $b$:

\[(a,b)=g_{ik} a^{i} b^{k} .\]

In other work of the author [6] relativistic - invariant function of Hamilton was entered by
relation: 

\begin{equation}\label{GrindEQ__6_}
H(x,P)=\sqrt{(P,P)} -q\Phi .
\end{equation}
In this work, we will give another, more flexible, definition of invariant Hamilton function of particles in the scalar field, corresponding its zero normalisation [11]:
\begin{equation} \label{GrindEQ__7_}
H(x,P)=\frac{1}{2} [m_{*}^{-1} (P,P)-m_{*} ]=0,
\end{equation}
where $m_{*} (\Phi )$ is while arbitrary scalar function. From definition \eqref{GrindEQ__7_} follows, that the vector of the generalized momentum is time-like:

\begin{equation} \label{GrindEQ__8_}
(P,P)=m_{*}^{2} .
\end{equation}
Let's note useful for further the relation, a being consequence \eqref{GrindEQ__3_}, \eqref{GrindEQ__7_} and \eqref{GrindEQ__8_}:

\begin{equation} \label{GrindEQ__9_}
[H,P^{k} ]=\nabla ^{k} m_{*} \equiv g^{ik} \partial _{i} m_{*} .
\end{equation}

\noindent From the canonical equations \eqref{GrindEQ__1_} we are obtained connection between the generalized momentum and a vector of speed of a particle:

\begin{equation} \label{GrindEQ__10_}
u^{i} \equiv \frac{dx^{i} }{ds} =m_{*}^{-1} P^{i} \Rightarrow P^{i} =m_{*} u^{i} ,
\end{equation}
which is satisfy by normalisation relation  automatically:

\begin{equation} \label{GrindEQ__11_}
(u,u)=1.
\end{equation}

\subsection{The motions equations in a Lagrange formulation}

From second group of canonical equations \eqref{GrindEQ__1_} we obtained motions equations in the Lagrange formulation:

\begin{equation} \label{GrindEQ__12_}
\frac{d^{2} x^{i} }{ds^{2} } +\Gamma _{jk}^{i} \frac{dx^{j} }{ds} \frac{dx^{k} }{ds} =\partial _{,k} \ln |m_{*} |{\rm {\mathcal P}}^{ik} ,
\end{equation}
 where:

\begin{equation} \label{GrindEQ__13_}
{\rm {\mathcal P}}^{ik} ={\rm {\mathcal P}}^{ki} =g^{ik} -u^{i} u^{k}
\end{equation}
 is the tenmsor of orthogonal projection on direction $u$, thus, what:

\begin{equation} \label{GrindEQ__14_}
{\rm {\mathcal P}}^{ik} u_{k} \equiv 0;\quad {\rm {\mathcal P}}^{ik} g_{ik} \equiv 3.
\end{equation}
From this relations and Lagrange equations \eqref{GrindEQ__12_} follows strict consequention of the orthogonality velocity vector and acseleration vector:

\begin{equation} \label{GrindEQ__15_}
g_{ik} u^{i} \frac{du^{k} }{ds} \equiv 0.
\end{equation}
From relations \eqref{GrindEQ__8_}, \eqref{GrindEQ__10_}, and Lagrange equations \eqref{GrindEQ__12_} also, is followed, what scalar $\varphi $ have meaning \textit{ of the effective inert mass of particle, $m_{*} $, in a scalar field}:

\begin{equation} \label{GrindEQ__16_}
\varphi =m_{*} .
\end{equation}
 Let's notice, that to the specified choice of Hamilton function there corresponds the following function of action:

\noindent

\begin{equation} \label{GrindEQ__17_}
S=\int  m_{*} ds.
\end{equation}

\subsection{Integrals of the motion}
Let's find now conditions of existence of linear integral of the canonical equations of motion \eqref{GrindEQ__1_}, for what we will calculate a total derivative on canonical parametre from scalar product $(\xi ,P)$. Using the canonical equations of motion \eqref{GrindEQ__1_}, a relation of a normalisation \eqref{GrindEQ__8_}, and also connection of the generalised momentum with kinematic \eqref{GrindEQ__10_}, we will find:

\begin{equation} \label{GrindEQ__18_}
\frac{d(\xi ,P)}{ds} =\frac{1}{m_{*} } P^{i} P^{k} \mathop{L}\limits_{\xi } g_{ik} +\mathop{L}\limits_{\xi } m_{*} ,
\end{equation}
 where $\mathop{L}\limits_{\xi } $ is Lee derivation on direction $\xi $ \footnote{See for example [12].}. Believing further

\begin{equation} \label{GrindEQ__19_}
\frac{d(\xi ,P)}{ds} =0\Leftrightarrow (\xi ,P)={\kern 1pt} Const{\kern 1pt} ,
\end{equation}
taking into account arbitrariness of a vector of the generalised momentum and a relation of a normalisation for it, we will receive conditions of performance of this equality:

\begin{equation} \label{GrindEQ__20_}
\mathop{L}\limits_{\xi } g_{ik} =\rho g_{ik} \Rightarrow \rho =-\mathop{L}\limits_{\xi } \ln |m_{*} |.
\end{equation}
Substituting this result back in a relation \eqref{GrindEQ__18_}, we will receive definitively necessary and sufficient conditions of existence of linear integral of the initial equations of movement (see, for example, [14]):

\begin{equation} \label{GrindEQ__21_}
\mathop{L}\limits_{\xi } m_{*} g_{ik} =0.
\end{equation}
Thus,\textit{ that there was a linear integral of the canonical equations of motions \eqref{GrindEQ__1_} it is necessary and enough that the conformally corresponding space with metrics $m_{*} g_{ik} $ supposed group of motions with vector Killing }$\xi $. We will notice, that linear integrals \eqref{GrindEQ__19_} make sense a full momentum (at spatially-like vector $\xi $) or full energy (at is time-like vector $\xi $).

\subsection{The choice of mass function}

There the question arises on a choice of function $m_{*} (\Phi )$. Not concretising while this function, we will note following important circumstance. We will consider static fields $g_{ik} $ and $\Phi $, supposing time-like vector Killing $\xi ^{i} =\delta _{4}^{i} $, when full energy of a particle, $P_{4} $, is conserved.

Let's consider further frame of reference, in which $g_{\alpha 4} =0$ so coordinate $x^{4} $ coincides with world time $t$. Then from connection relations between a vector of kinematic speed $u^{i} $ and a vector of a total momentum of particle $P_{i} $ \eqref{GrindEQ__10_} follows:

\begin{equation} \label{GrindEQ__22_}
P_{4} ds=m_{*} dt,
\end{equation}
where $P_{4} =E_{0} ={\kern 1pt} Const{\kern 1pt} >0$ is full energy of charged particle. Therefore, if we wish to keep identical orientation of world and own time (i.e., $dt/ds>0$), it is necessary to choose such mass function which always would remain non-negative:

\begin{equation} \label{GrindEQ__23_}
m_{*} >0.
\end{equation}
Apparently, for example, from equations Lagrange \eqref{GrindEQ__12_}, this function is convenient for choosing so that:

\noindent

\begin{equation} \label{GrindEQ__24_}
m_{*} (\Phi )=|m_{*} (\Phi )|\ge 0.
\end{equation}
Further, on the one hand, for absence of a scalar field, more precisely, in a constant scalar field, mass function should pass in rest-mass of a particle, $m\ge 0$. On the other hand, Lagrange equations \eqref{GrindEQ__12_} in case of a weak scalar field should pass in the classical equations of motion in a scalar field.

\noindent Thus, proceeding from a accordance principle, we should have:

\begin{equation} \label{GrindEQ__25_}
m_{*} (0)=m;\quad \quad (m_{*} )_{,k} |_{\Phi =0} =q\Phi _{,k} ,
\end{equation}
 where $q$ is some fundamenta constance -- \textit{ scalar charge of particle}. Conditions \eqref{GrindEQ__25_} mean, that at small values of scalar potential $\Phi $ function $m_{*} (\Phi )$ should have kind decomposition:

\begin{equation} \label{GrindEQ__26_}
m_{*} (\Phi ){\rm \simeq }m(1+\frac{q\Phi }{m} +..).
\end{equation}
To this condition corresponds and the linear function used in quoted works $m_{*} (\Phi )=|m+q\Phi |$.

It is possible to offer and other, more radical approach, which thus does not contradict relations \eqref{GrindEQ__25_} and \eqref{GrindEQ__26_}, believing that all inert mass of particles arises owing to interaction with a scalar field: 
\begin{equation}\label{GrindEQ__27_} \varphi (\Phi )\equiv m_{*} =|q\Phi |.\end{equation}
Then under rest-mass of a particle, $m_{0} $, it is necessary to understand its mass \eqref{GrindEQ__27_} at a modern stage of evolution of the Universe, to which there corresponds scalar potential $\Phi _{0} $: $m_{0} =m_{*} (\Phi (t_{0} ))$. To such choice of function $\phi (\Phi )$ there corresponds action function:

\begin{equation} \label{GrindEQ__28_}
S=\int  |q\Phi |ds.
\end{equation}
This choice answers also to aesthetic criteria as in this case function of Hamilton \eqref{GrindEQ__7_} does not depend on rest-mass. On the other hand it is visible, that at a choice of function $\varphi (\Phi )$ in the form of \eqref{GrindEQ__27_} Lagrange equations  \eqref{GrindEQ__12_} become symmetric concerning replacement $\Phi \to -\Phi $ and under condition of $q\ne 0$ do not depend at all obviously on a scalar charge:

\begin{equation} \label{GrindEQ__29_}
\frac{d^{2} x^{i} }{ds^{2} } +\Gamma _{jk}^{i} \frac{dx^{j} }{ds} \frac{dx^{k} }{ds} =(\ln |\Phi |)_{,k} {\rm {\mathcal P}}^{ik} ,
\end{equation}
At a conclusion of the equations \eqref{GrindEQ__29_} we have considered differential identity: $d|y|={y\mathord{\left/ {\vphantom {y |y|}} \right. \kern-\nulldelimiterspace} |y|} dy$. However, by means of a relation of a normalisation \eqref{GrindEQ__8_} dependence of decisions on a scalar charge nevertheless remains and at such choice of function of weight in the form of dependence of energy on impulse $P_{4} (P^{2} )$.

\subsection{One-dimensional motion}

 Let's consider the following problem. Let in space Minkovsky there is the static scalar field which potential depends only on one coordinate, $x^{1} =x$, and let for simplicity $m_{*} =|q\Phi |=|x|$. Thus, there are 3 vectors Killing - one time-like and two spaselike:

\def\stackunder#1#2{\mathrel{\mathop{#2}\limits_{#1}}}
$$\stackunder{1}{\xi}^i=\delta^i_4;\:\stackunder{2}{\xi}^i=\delta^i_2;\:\stackunder{3}{\xi}^i=\delta^i_3. $$

According to these Killing vectors there are three linear integrals of motion: 
\begin{equation}\label{GrindEQ__30_} P_{2} =P_{2}^{0} ={\kern 1pt} Const{\kern 1pt} ;\; P_{3} =P_{3}^{0} ={\kern 1pt} Const{\kern 1pt} ;\; P_{4} =P_{4}^{0} ={\kern 1pt} Const{\kern 1pt} .\end{equation}
 Let for simplicity $P_{2} =P_{3} =0$. Then taking into account a relation of a normalisation \eqref{GrindEQ__8_} we will receive from the initial equations of motions one not trivial:

\begin{equation} \label{GrindEQ__31_}
\frac{dx}{dt} =\mp \frac{\sqrt{P_{4}^{2} -x^{2} } }{P_{4} } .
\end{equation}
 Let's put for simplicity $P_{4} =1$, $x(0)=1/2$. Then in the equation \eqref{GrindEQ__31_} it is necessary to choose a negative sign in the right part. The solve of this equation is:

\begin{equation} \label{GrindEQ__32_}
x=\cos (t+\pi /3);
\end{equation}
 -- and it describes harmonious oscilations in world time $t$; thus generalised momentum, $P_{1} $ also is harmonious function of world time:

\begin{equation} \label{GrindEQ__33_}
P_{1} =\sin (t+\pi /3).
\end{equation}
However, coordinates of 4 measured vectors of kinematic speed of a particle, $u^{i} $

\[u^{1} =\frac{dx}{ds} \equiv -P_{1} /\phi =-\tan (t+\pi /3)\]
Undergo ruptures of 2nd sort during the moments of time $t=\pi /6+\pi k$, in which $x=0$. It testifies or to connection rupture between coordinates and own time during the specified moments of world time, or about necessity of redefinition of own time for scalar charged particles. For a kinematic momentum of a particle, $p^{i} $ if we enter it as

\begin{equation} \label{GrindEQ__34_}
p^{i} =m_{*} \frac{dx^{i} }{ds} \equiv P^{i} ,
\end{equation}
such problem does not arise, as well as for three-dimensional speed $v^{\alpha } =u^{\alpha } /u^{4} $.

\noindent Let's notice, that, actually, only these continuous sizes and are physically measurable.

Nevertheless, the specified example shows, that it is necessary to spend accurately calculations for scalar charged particles. Further we will be yet we will concretise a normalisation of effective weight, believing only executed a relation \eqref{GrindEQ__24_}.


\begin{thebibliography}{10}
\bibitem{1}  D. Galtsov è E. Davidov. // In book::  Quantum theory and cosmology. -- Sankt-Peterburg: Freedman Laboratory Publ. -  2009. -- p. 25-44.
\bibitem{2}   Yu.G. Ignatyev, R.F. Miftakhov. // Proc.  II-nd Russian Summer School «Actual Theoretical Problems of Gravitation and Cosmology». -- 2009. -- Kazan, Russia, «Foliant» publ.. -- p.  76-83.
\bibitem{3}   V.M. Zhuravlev, R.R. Abbyazov. //\textit{ Gravitation and Cosmology}, bf 16, No 1, 50, (2010).
\bibitem{4}  V.M. Zhuravlev. // ZETP, \textbf{ 120}, 1042, (2001).

\bibitem{5}  Yu. Ignat'ev, R. Miftakhov. // Gravitation \& Cosmology. -- 2006. - \textbf{ 12}, No 2-3. -- p.179-184;ñì. òàêæå: Yu. Ignat'ev, R. Miftakhov. // arXiv:1011.5774[gr-qc].
\bibitem{6}  Yu.G. Ignatyev. // Russian Physics Journal. -  1982. - ¹ 4. -- Ñ. 92-93.

\bibitem{7}  Yu.G. Ignatyev. // Russian Physics Journal. -  1983. - ¹ 8. -- Ñ. 15-19.

\bibitem{8}  Yu.G. Ignatyev. // Russian Physics Journal. -  1983. - ¹ 8. -- Ñ. 19-23.

\bibitem{9}  Yu.G. Ignatyev. // Russian Physics Journal. -  1983. - ¹ 12. -- Ñ. 9-13.

\bibitem{10}   G.G. Ivanov. // Russian Physics Journal. -  1983. - ¹ 1. -- Ñ. 32-36.

\bibitem{11}  Yu.G. Ignatev and A.A. Popov. // Actrophysics and Space Science. -- 1990. --\textbf{ 163. -}  p. 153-174; Yu.G. Ignatyev, A.A. Popov. // arXiv:1101.4303v1 [gr-qc].

\bibitem{12}  A.Z. Petrov. New methods in General Theory Relativity. -- Moskow.: Nauka publ. -- 1966. -- 496 p.

\bibitem{13}  Yu.G. Ignatev. The Relativistic Kinetic Theory of the Nonequilibrium Processes in a Gravitational Fields. -- -- 2009. -- Kazan, Russia, «Foliant» publ.. -- 2010. -- 506 p.;  http://rgs.vniims.ru/books/const.pdf.

\bibitem{14}  Yu.G. Ignatev, R.R. Kuzeev. Ukr. Fiz. Journal., \textbf{ 29}, 1021, (1984).
\end{thebibliography}
\end{document}